\documentclass[aps,prd,twocolumn,showkeys,nofootinbib]{revtex4}
\usepackage[dvipsnames]{xcolor}
\usepackage{color}
\pdfoutput=1
\usepackage{epsfig}
\usepackage{amsmath}
\usepackage{graphicx}
 \usepackage{multirow}
\newcommand{\beq}{\begin{equation}}
\newcommand{\eeq}{\end{equation}}
\newcommand{\bqa}{\begin{eqnarray}}
\newcommand{\eqa}{\end{eqnarray}}

\begin{document}

\title{Interpretation of the newly observed $\Omega_c^0$ resonances}
\author{Wei Wang$^2$, Rui-Lin Zhu$^1$~\footnote{Corresponding author, Email:rlzhu@njnu.edu.cn} }
\affiliation{
$^1$ Department of Physics and Institute of Theoretical Physics,
Nanjing Normal University, Nanjing, Jiangsu 210023, China \\
$^2$ INPAC, Shanghai Key Laboratory for Particle Physics and Cosmology, School of Physics and Astronomy, Shanghai Jiao Tong University, Shanghai, 200240,   China}

%\date{\today}
\begin{abstract}
We study the  charmed and bottomed doubly strange  baryons within   the heavy-quark-light-diquark framework.  The two strange quarks are assumed to lie in  $S$ wave and thus their total spin  is 1. We calculate  the mass spectra  of the $S$ and $P$ wave orbitally excited states and  find the   $\Omega_c^0 (2695)$ and $\Omega_c^0 (2770)$ fit well as the $S$ wave states of  charmed doubly strange baryons.   The five newly  $\Omega_c^0(X)$ resonances observed by the LHCb Collaboration, i.e. $\Omega_c^0(3000)$, $\Omega_c^0(3050)$, $\Omega_c^0(3066)$, $\Omega_c^0(3090)$, and $\Omega_c^0(3119)$, can be interpreted as the $P$ wave orbitally excited states. In   heavy quark effective theory,  we analyze their decays into the $\Xi^+_c K^-$ and $\Xi_c'^+ K^-$, and point out that  decays of  the five P-wave $\Omega_c^0$ states into the $\Xi_c^+ K^-$ and $\Xi_c'^+K^-$ are suppressed by either  heavy quark symmetry or   phase space. The narrowness of the five newly  observed $\Omega_c^0(X)$ states   can then be naturally  interpreted  with heavy quark symmetry.


\end{abstract}

\maketitle

\section*{Introduction}

The hadron spectroscopy plays an important role in understanding the fundamental  theory of strong interactions, i.e. the Quantum Chromodynamics (QCD). In the naive quark model, the mesons are bound states of a quark-antiquark pair while the baryons are composed of three quarks. However, the structure of hadrons is more complicated than the description  in the naive quark model. There might be  hybrids, glueballs, and multiquark states, which are also allowed under the principle of color confinement. Take the exotic baryon states as example, the LHCb Collaboration have observed two pentaquark candidates $P_c(4380)$ and $P_c(4450)$ in $\Lambda^0_b\to J/\psi K^- p$ decays~\cite{Aaij:2015tga},  which also have been analyzed in $\Lambda^0_b\to J/\psi \pi^- p$ decays~\cite{Aaij:2016ymb}. The general studies of hadron inner structures will enhance our knowledge on the properties of QCD  color confinement.

The charmed doubly strange baryon $\Omega_c^0(2695)$ with isospin and spin-parity $I(J^{P})=0(\frac{1}{2}^+)$ was first observed in the hyperon beam experiment WA62~\cite{Biagi:1984mu}. Later it was confirmed  in the electron-positron collider experiment~\cite{Albrecht:1992xa}  and the photon beam experiment~\cite{Frabetti:1992bm}. The excited state $\Omega_c^0(2770)$ with $I(J^{P})=0(\frac{3}{2}^+)$ was first observed in the radiative decay $\Omega_c^0(2770)\to \Omega_c^0(2695)+\gamma $  by the BaBar  Collaboration~\cite{Aubert:2006je}, and then confirmed by the Belle
Collaboration~\cite{Solovieva:2008fw}.

Using a sample of $pp$ collision data corresponding to an integrated luminosity of 3.3$fb^{-1}$, the LHCb Collaboration has  recently observed
five new narrow excited $\Omega_c^0(X)$ states  in the $\Xi_c^+ K^-$ invariant mass spectrum~\cite{Aaij:2017nav}. They have determined the masses and decay widths of the five new $\Omega_c^0(X)$ states~\cite{Aaij:2017nav}   and the results  are collected in Table~\ref{table0}.

%%%%%%%%%%%%%%%%%%%
\begin{table}
\begin{center}
\caption{Masses and widths  (MeV) of the $\Omega_c^0(X)$ baryons observed by the LHCb Collaboration. The first uncertainty is statistical and the second one is systematic, and the third uncertainty in masses of $\Omega_c^0(X)$ baryons is from the $\Xi_c^+$ mass.
}\label{table0}
\begin{tabular}{c|ccccccccccc}
 \hline\hline
state  & mass  & width   \\
 \hline
 $\Omega_c(3000)$ & $3000.4\pm0.2\pm0.1_{-0.5}^{+0.3}$ & $4.5\pm 0.6\pm0.3$  \\
 \hline
 $\Omega_c^0(3050)$ & $3050.2\pm0.1\pm0.1_{-0.5}^{+0.3}$ & $0.8\pm 0.2\pm0.1$   \\
 \hline
  $\Omega_c^0(3066)$ &$3065.6\pm0.1\pm0.3_{-0.5}^{+0.3}$ ~& $3.5\pm 0.4\pm0.2$    \\
     \hline
    $\Omega_c^0(3090)$ & $3090.2\pm0.3\pm0.5_{-0.5}^{+0.3}$ ~& $8.7\pm 1.0\pm0.8$  \\
 \hline
  $\Omega_c^0(3119)$ & $3119.1\pm0.3\pm0.9_{-0.5}^{+0.3}$ ~& $1.1\pm 0.8\pm0.4$\\
    \hline\hline
\end{tabular}
\end{center}
\end{table}
%%%%%%%%%%%%%%%%%%%

After these discoveries, it is natural to ask ourselves three questions: 1) Why are there so small mass differences among
these five new states? 2)What are the spin-parities for these five new states? 3)Why are the decay widths so narrow for these five new states?

Investigating their mass spectra and decay properties will answer these questions accordingly.
In theoretical aspects, there are already some attempts  to interpret of the newly observed $\Omega_c^0(X)$ resonances. Agaev et al. proposed to assign $\Omega_c^0(3066)$ and $\Omega_c^0(3119)$  states as the first radially excited $(2S, \frac{1}{2}^+)$ and $(2S, \frac{3}{2}^+)$ charmed baryons in QCD sum rules~\cite{Agaev:2017jyt}. Chen et al. analyzed the newly  $\Omega_c^0(X)$  states with different spins and obtained the related decay widths into $\Xi_c^+ K^-$, $\Xi_c'^+ K^-$ and $\Xi_c^{*+} K^-$  in QCD sum rules~\cite{Chen:2017sci}. Karliner et al. proposed to assign the newly  $\Omega_c^0(X)$  states  as bound states of a charm quark and a $P$ wave $ss$-diquark~\cite{Karliner:2017kfm}. Wang et al. studied the strong and radiative decays of the $\Omega_c^0(X)$  states in a constituent quark model~\cite{Wang:2017hej}. Besides,  Yang et al. proposed to assign some of the newly  $\Omega_c^0(X)$  states  as the possible pentaquark states~\cite{Yang:2017rpg}.

In this paper, we will interpret the five new  observed $\Omega_c^0(X)$  states as the $P$ wave orbitally excited states of  charmed doubly strange baryons in the heavy-quark-light-diquark picture. The spectra of the bottom partners of $\Omega_c^0(X)$  states will be also predicted. In the end, the decay properties of  charmed doubly strange baryons will be  discussed in the heavy quark  effective theory.

\section*{Interpretation of the newly observed $\Omega_c^0$ resonances}
%%%%%%%%%%%%%%%%%%%%%%%%%%%%%%%%%%%%%%%%%%%%%%%%%%
The notion of diquark is as old as the quark model where Gell-Mann mentioned the possibility of diquarks in the original paper on quarks~\cite{GellMann:1964nj}.
According to the color SU(3) group,  the color configuration of a diquark can be represented either by an antitriplet or sextet in the decomposition of  $\mathbf{3}\otimes \mathbf{3}=\mathbf{\bar{3}}\oplus \mathbf{6}$. The binding of the $q_1\bar{q}_2$ or $q_1q_2$ system depends solely on the quadratic Casimir $C_2(R)$ of the product color representation
R to which the quarks couple according to the discriminator
$I=\frac{1}{2}(C_2(R)-C_2(R_1)-C_2(R_2))$, where $R_i$ denotes the color representations of two quarks \cite{Brodsky:2014xia}. The discriminators are then determined as $I=\frac{1}{6}(-8,-4,+2,+1)$ for $R = (\mathbf{1},\mathbf{\bar{3}},\mathbf{6},\mathbf{8})$, respectively. The interaction force becomes attractive when
the discriminator is negative, which is
somewhat analogous to the Coulomb force in QED. Thus, the only color attractive configuration of $q_1\bar{q}_2$ is in the color-singlet $\mathbf{1}$, whereas the color attractive configuration of $q_1q_2$ is in the color antitriplet $\mathbf{\bar{3}}$.  The attractive force strength in the color antitriplet diquark is half of that in  the color singlet quark-antiquark pair in the one-gluon-exchange model.
Thus two   quarks in the color antitriplet $\mathbf{\bar{3}}$ have a large possibility  to bind into a  diquark ~\cite{Brodsky:2014xia,Jaffe:2003sg,Jaffe:2004ph}, and thus a baryon can be treated as a quark-diquark system.

In the $css$ system,
two strange quarks can form a light diquark system, while the charm and strange quarks may also form a $cs$ diquark.  The strength of the attractive force between two quarks is reflected  by a coupling constant as given below.  A fit of the experimental data  have indicated that the coupling constant for the two strange quarks is much larger than that for the $cs$ system, for instance, $\kappa_{ss}=72$MeV and $\kappa_{cs}\simeq (24-25)$MeV~\cite{Maiani:2004vq,Ali:2011ug,Hambrock:2011qga}. Following this scheme, we will treat the charmed doubly strange baryons as
heavy-quark-light-diquark bound states  in order  to explain the newly observed five narrow  $\Omega_c^0(X)$ states.

The   wave function of the charmed doubly strange baryon is composed of  four parts,  coordinate-space, color, flavor, and spin subspaces~\cite{He:2016xvd}
 \begin{eqnarray}
\Psi(c,s,s)&=&\psi(x_1,x_2,x_3)\otimes
\chi_{123}\otimes f_{123}\otimes s_1s_2s_3\,,
\end{eqnarray}
where we denote numbers 1, 2, 3 to charm and two strange quarks respectively; $\psi(x_i)$, $\chi$, $f$, and $s_i$ denote the coordinate-space, color, flavor, and spin wave functions, respectively. The total wave function should satisfy  the Pauli exclusion  principle   when we interchange the two strange quarks.   We will restrict ourselves to the ground state of the diquark, namely the coordinate-space wave function  is in the $S$-wave  with $L=0$, and thus symmetric.  The color wave function is  anti-symmetrical because the baryon system is in the color singlet.
The flavor wave function is also symmetrical to the interchange of  the two strange quarks. Thus the spin wave function should be also symmetrical, i.e. the spin of two strange quarks should be 1 in the charmed doubly strange baryon.

The charmed doubly strange baryons are composed of a charm quark and two strange quarks. We
assume the two strange quarks form a diquark  $\delta=ss$, which along with the charm quark
make it true for the stable spectra of the $\Omega_c^0(X)$  system. The baryons mass splitting $\Delta M$
can be estimated as~\cite{Jaffe:2003sg,Maiani:2004vq}
\begin{eqnarray}
\Delta M&=&2(\kappa_{c s})_{\bar{3}}(\mathbf{S}_c\cdot \mathbf{S}_{\delta})+2(\kappa_{s s})_{\bar{3}}(\mathbf{S}_s\cdot \mathbf{S}_{s})\nonumber\\
&&+2A_c(\mathbf{S}_c\cdot \mathbf{L})+2A_\delta(\mathbf{S}_\delta\cdot \mathbf{L})
\nonumber\\
&&+B\frac{L(L+1)}{2},
\label{eq:definition-hamiltonian}
\end{eqnarray}
where the  first two terms are spin-spin interaction between the diquark and charm quarks and  inside the diquark. The third and fourth terms are the spin-orbital interactions. The fifth term is the pure orbital interactions. The $\mathbf{S}_{\delta}$  corresponds to the spin operator  of diquark. The spin operators  of strange quark and  charm quark  are given by  $\mathbf{S}_{s}$ and $\mathbf{S}_{c}$, respectively.   The coefficients  $(\kappa_{q_1 q_2})_{\bar{3}}$ are the spin-spin couplings for two quarks in color antitriplet, respectively.

Unlike the case in the $\Omega^-$ where the total angular momentum  $J$ is $3/2$ with $L=0$, the $S$ wave states of the $\Omega_c^0$ system have two states where the total angular momentum  $J$ can be either $1/2$ or $3/2$.

\begin{eqnarray}
|L=0,\frac{1}{2}_J\rangle&=&|\frac{1}{2}_c,1_{\delta};\frac{1}{2}_{c\delta};L=0;\frac{1}{2}_J\rangle
\nonumber\\&=&\frac{\sqrt{2}}{\sqrt{3}}(\downarrow)_c
(\uparrow)_s (\uparrow)_s,\\
|L=0,\frac{3}{2}_J\rangle&=&|\frac{1}{2}_c,1_{\delta};\frac{3}{2}_{c\delta};L=0;\frac{3}{2}_J\rangle\nonumber\\
&=&(\uparrow)_c
(\uparrow)_s (\uparrow)_s,
 \label{eq:definition1}
\end{eqnarray}
where $|S_c,S_{\delta};S_{c\delta};L=0;N_J\rangle$ stands for the baryon; the $S_\delta$ and $S_c$ denote  the spin  of the diquark $[ss]$ and the charm quark, respectively, and the $N_J$ denotes the total angular momentum of the baryon.

There are five $P$ wave states of $\Omega_c^0$ system  with $L=1$ and negative parity
\begin{eqnarray}
|L=1,\frac{1}{2}_J\rangle_1&=&|\frac{1}{2}_c,1_{\delta};\frac{1}{2}_{c\delta};L=1;\frac{1}{2}_J\rangle,\label{eq:five_states_diquark1}\\
|L=1,\frac{1}{2}_J\rangle_2&=&|\frac{1}{2}_c,1_{\delta};\frac{3}{2}_{c\delta};L=1;\frac{1}{2}_J\rangle,
\end{eqnarray}
\begin{eqnarray}
|L=1,\frac{3}{2}_J\rangle_1&=&|\frac{1}{2}_c,1_{\delta};\frac{1}{2}_{c\delta};L=1;\frac{3}{2}_J\rangle,\\
|L=1,\frac{3}{2}_J\rangle_2&=&|\frac{1}{2}_c,1_{\delta};\frac{3}{2}_{c\delta};L=1;\frac{3}{2}_J\rangle,\\
|L=1,\frac{5}{2}_J\rangle&=&|\frac{1}{2}_c,1_{\delta};\frac{3}{2}_{c\delta};L=1;\frac{5}{2}_J\rangle.\label{eq:five_states_diquark}
 \label{eq:definition2}
\end{eqnarray}

There are some simple relations among the $S$ and $P$ wave states of $\Omega_c^0$ system when using the
mass splitting formulae.  Their relations are
\begin{eqnarray}
M_{|L=0,\frac{3}{2}_J\rangle}&=&M_{|L=0,\frac{1}{2}_J\rangle}+3(\kappa_{c s})_{\bar{3}}, \label{eq:mass1}\\
M_{|L=1,\frac{1}{2}_J\rangle_1}&=&M_{|L=0,\frac{1}{2}_J\rangle}-2A_c+B,\\
M_{|L=1,\frac{1}{2}_J\rangle_2}&=&M_{|L=0,\frac{1}{2}_J\rangle}+3(\kappa_{c s})_{\bar{3}}-5A_c+B,\\
M_{|L=1,\frac{3}{2}_J\rangle_1}&=&M_{|L=0,\frac{1}{2}_J\rangle}+A_c+B,\\
M_{|L=1,\frac{3}{2}_J\rangle_2}&=&M_{|L=0,\frac{1}{2}_J\rangle}+3(\kappa_{c s})_{\bar{3}}-2A_c+B,\\
M_{|L=1,\frac{5}{2}_J\rangle}&=&M_{|L=0,\frac{1}{2}_J\rangle}+3(\kappa_{c s})_{\bar{3}}+3A_c+B,
 \label{eq:mass}
\end{eqnarray}
where we simply assume $A_\delta=A_c$.

For convenience, we write the possible states into the corresponding form $|n^{2S+1}L_J\rangle$, i.e.
$|1^{2}S_{\frac{1}{2}}\rangle=|L=0,\frac{1}{2}_J\rangle$, $|1^{4}S_{\frac{3}{2}}\rangle=|L=0,\frac{3}{2}_J\rangle$,
$|1^{2}P_{\frac{1}{2}}\rangle=|L=1,\frac{1}{2}_J\rangle_1$, $|1^{4}P_{\frac{1}{2}}\rangle=|L=1,\frac{1}{2}_J\rangle_2$,
$|1^{2}P_{\frac{3}{2}}\rangle=|L=1,\frac{3}{2}_J\rangle_1$, $|1^{4}P_{\frac{3}{2}}\rangle=|L=1,\frac{3}{2}_J\rangle_2$,
and  $|1^{4}P_{\frac{5}{2}}\rangle=|L=1,\frac{5}{2}_J\rangle$.  Assuming the $\Omega_c^0(2695)$ is the ground state with $1^{2}S_{\frac{1}{2}}$ and then  the $\Omega_c^0(2695)$ is the lightest state, the mass
spectra of the $S$ and $P$ wave states of $\Omega_c^0(X)$ baryons can be obtained from the relations
in Eqs.~(\ref{eq:mass1}-\ref{eq:mass}).

The coupling constants
in Eq.~\ref{eq:definition-hamiltonian} are described in detail in Refs.~\cite{Jaffe:2004ph,Maiani:2004vq,Ali:2009es,Zhu:2015bba,Wang:2016tsi,Zhu:2016arf}. In order to give more information of the coupling constants, we extract the coupling constants from the baryon mass relations
~\cite{Jaffe:2004ph}
\begin{eqnarray}
(\kappa_{c s})_{\bar{3}}&=&2 K(c,\{u,s\})-K(c,\{u,d\}),\\
K(c,\{u,d\})&=&\frac{1}{3}(m_{\Sigma_c^{*+}}-m_{\Sigma_c^{+}}),\\
K(c,\{u,s\})&=&\frac{1}{6}(2m_{\Xi_c^{*0}}-m_{\Omega_c^0}-m_{\Sigma_c^{+}}).
\label{eq:coupling}
\end{eqnarray}
Inputting the related charmed baryon masses~\cite{Olive:2016xmw}, i.e. $m_{\Xi_c^{*0}}=(2645.9\pm0.5)$MeV, $m_{\Omega_c^0}=(2695.2\pm1.7)$MeV, $m_{\Sigma_c^{+}}=(2452.9\pm0.4)$MeV, and $m_{\Sigma_c^{*+}}=(2517.5\pm2.3)$MeV,
the value of the coupling constant $(\kappa_{c s})_{\bar{3}}$ can be extracted
as $(\kappa_{c s})_{\bar{3}}=(26\pm1.5)$MeV.

The parameters $A_c$ and $B$ which describe the orbital couplings of the excited states can be estimated by the comparison with the observed spin-orbitally splitting
in the $\Xi_c^0(X)$ states. We have the estimation
\begin{eqnarray}
-2A_c+B&\simeq&m_{\Xi_c^{0}(\frac{1}{2}^-)}-m_{\Xi_c^{0}(\frac{1}{2}^+)},\\
A_c+B&\simeq&m_{\Xi_c^{0}(\frac{3}{2}^-)}-m_{\Xi_c^{0}(\frac{1}{2}^+)}.
\label{eq:coupling}
\end{eqnarray}
Inputting the related charmed baryon masses~\cite{Olive:2016xmw}, i.e. $m_{\Xi_c^{0}(\frac{1}{2}^+)}=(2470.85^{+0.28}_{-0.40})$MeV, $m_{\Xi_c^{0}(\frac{1}{2}^-)}=(2791.9\pm3.3)$MeV, and $m_{\Xi_c^{0}(\frac{3}{2}^-)}=(2819.6\pm1.2)$MeV,
the value of the coupling constants can be extracted
as $A_c(\Omega_c)=(9\pm1.5)$MeV and $B(\Omega_c)=(340\pm2)$MeV.

Considering the uncertainties of the inputting parameters, the mass
spectra of the $S$ and $P$ wave states of $\Omega_c^0(X)$ baryons are given in Tab.~\ref{table1}. In this table, the   assignment  of  $\Omega_c^0$ baryons to  $|n^{2S+1}L_J\rangle$ is by no means conclusive. For instance,  the $\Omega_c(2695)$ has been assigned as the ground state only due to the fact there is no other lower state that has been established on the experimental side. In Tab.~\ref{table1} we also list the experimental data and other theoretical predictions. Most of them are based on the potential model, QCD sum rules, and Lattice QCD simulation. Besides, some excited states of $\Omega_c^0(X)$ baryons are also predicted from meson-baryon unitarization starting from a lowest order potential in Refs.~\cite{JimenezTejero:2009vq,Romanets:2012hm}, where  the existence of a bound state
at 2959 MeV, near the lowest threshold, and two resonances
placed at 2966 and 3117 MeV are predicted in this scheme.  The widths of the two resonances are calculated as $\Gamma(2966)=1.1$ MeV and $\Gamma(3117)=16$ MeV.

%%%%%%%%%%%%%%%%%%%%%%%%%%%%%%%
\begin{table*}[ht]
\begin{center}
\caption{The mass spectra (MeV) of $\Omega_c^0(X)$ baryons. The uncertainties of the experimental measurements are squared averages of those from the statistical and systematic, and the $\Xi_c^+$ mass.
}
\begin{tabular}{c|ccccccccccc}
 \hline\hline
$n^{2S+1}L_J$ & This work ~~& Exp.~\cite{Aaij:2017nav,Olive:2016xmw}~\footnote{The following assignment  of  $\Omega_c^0$ baryons to  $|n^{2S+1}L_J\rangle$ is by no means conclusive. For instance,  the $\Omega_c(2695)$ has been assigned as the ground state only due to the fact there is no other lower state that has been established on the experimental side. } & \cite{Shah:2016nxi}~ & \cite{Ebert:2011kk}~~  & \cite{Roberts:2007ni}~~
    &\cite{Yoshida:2015tia} ~~& \cite{Chen:2015kpa}~~& \cite{Valcarce:2008dr}~~&\cite{Yamaguchi:2014era}~~ & \cite{Bali:2015lka}~~\\
 \hline
 $1^{2}S_{\frac{1}{2}}$ & $2695.2\pm1.7$~~~& $2695.2\pm1.7$~~& 2695~~ & 2698~~ & 2718~~
    &2731~~ &-- &2699~~ &2648~~ & 2718~~\\
 \hline
 $1^{4}S_{\frac{3}{2}}$ & $2773\pm6$ ~~& $2765.9\pm2.0$~~& 2767~~ & 2768~~ & 2776~~
    &2779~~ & --&2767~~ &-- &-- \\
 \hline
  $1^{2}P_{\frac{1}{2}}$ & $3068\pm16$  ~& $3050.2\pm0.5$~~& 3011~~ &3055~~  & 2977~~
    &3030~~ & 3250~~&2980~~ & 2995~~& 3046~~\\
     \hline
    $1^{2}P_{\frac{3}{2}}$ & $3095\pm11$  ~~& $3090.2\pm0.8$ ~~& 2976~~ & 3054~~ & 2986~~
    &3033~~&3260~~ & 2980~~&3016~~ & 2986~~\\
 \hline
  $1^{4}P_{\frac{1}{2}}$ & $3017\pm7$ ~& $3000.4\pm0.5$ ~~& 3028~~ & 2966~~ & 2977~~
    &-- & --& 3035~~&-- &-- \\
 \hline
  $1^{4}P_{\frac{3}{2}}$ & $3044\pm5$ ~~& $3065.6\pm0.6$ ~~& 2993~~ & 3029~~ & 2959~~
    &-- &-- &-- & --&-- \\
 \hline
  $1^{4}P_{\frac{5}{2}}$ & $3140\pm13$ ~&$3119.1\pm1.1$ ~~& 2947~~ & 3051~~ & 3014~~
    &3057~~ &3320~~ &-- &-- &3014~~ \\
    \hline\hline
\end{tabular}\label{table1}
\end{center}
\end{table*}
%%%%%%%%%%%%%%%%%%%%%%%%%%%%%%%

The bottom partners of the $\Omega_c^0(X)$ baryons  can also be predicted.
Assuming the $\Omega_b^-(6046)$ with the mass $(6046\pm1.9)\mathrm{MeV}$ is the lightest state with $1^{2}S_{\frac{1}{2}}$, the spectra of $\Omega_b^-(X)$ baryons are very similar to that of $\Omega_c^0(X)$ baryons. Their masses and spin-parities are estimated as
\begin{eqnarray}
M_{\Omega_b}(1^{4}S_{\frac{3}{2}})&=&(6121\pm8)\mathrm{MeV},\nonumber\\
M_{\Omega_b}(1^{2}P_{\frac{1}{2}})&=&(6444\pm10)\mathrm{MeV},\nonumber\\
M_{\Omega_b}(1^{2}P_{\frac{3}{2}})&=&(6459\pm8)\mathrm{MeV},\nonumber\\
M_{\Omega_b}(1^{4}P_{\frac{1}{2}})&=&(6504\pm22)\mathrm{MeV},\nonumber\\
M_{\Omega_b}(1^{4}P_{\frac{3}{2}})&=&(6519\pm16)\mathrm{MeV},\nonumber\\
M_{\Omega_b}(1^{4}P_{\frac{5}{2}})&=&(6544\pm18)\mathrm{MeV},
 \label{eq:mass2}
\end{eqnarray}
where the parameters  are adopted as
$(\kappa_{b s})_{\bar{3}}=25\pm2$MeV, $A_b(\Omega_b)=5\pm2$MeV, and $B(\Omega_b)=408\pm4$MeV~\cite{Ali:2011ug,Hambrock:2011qga,Ali:2009es}. Since the observed spin-orbitally splitting in the $\Xi_b^-(X)$ states is limited, we only give the approximate error and will discuss the uncertainties of the coupling constants in future works. The mass splitting for the P-wave orbitally excited states is very small. Currently, only the ${\Omega_b}(1^{2}S_{\frac{1}{2}})$ has been observed~\cite{Olive:2016xmw}. The  S-wave orbitally excited state ${\Omega_b}(1^{4}S_{\frac{3}{2}})$ and the five P-wave orbitally excited states can be also reconstructed
by the electro-weak decay channel $\Omega^-_b(X) \to J/\psi+ \Omega^-$ with the sub-decays $ J/\psi\to\mu^+\mu^-(e^+e^-)$
and $\Omega^-\to \Lambda K^- (\Xi^0\pi^-)\to p \pi^- K^- ( p \pi^-\pi^0\pi^-)$. This can be examined in future.

\section*{Decays into $\Xi_cK$ and $\Xi_c^{'}K$}
In the heavy quark limit, the static heavy quark can only interact
with gluons via its chromoelectric charge, which leads to the heavy quark spin symmetry. In this heavy quark limit, the spin of the heavy quark and  the light degrees
of freedom $\mathbf{S}_\ell=\mathbf{J}-\mathbf{S}_Q$ with $Q=c,b$  is conserved, respectively. Thus some relations for the strong decays can be obtained.

In the heavy quark limit, the five P-wave baryonic states are given as
\begin{eqnarray}
 |  \frac{1}{2}_J\rangle_{1'} &\equiv& |\frac{1}{2}_c; {S_\ell=0}\rangle , \\
 | \frac{1}{2}_J\rangle_{2'} &\equiv&|\frac{1}{2}_c; {S_\ell=1}\rangle_1,
 \end{eqnarray}
 \begin{eqnarray}
 | \frac{3}{2}_J\rangle_{1'}&\equiv& |\frac{1}{2}_c; {S_\ell=1}\rangle_2,\\
 | \frac{3}{2}_J\rangle_{2'} &\equiv&|\frac{1}{2}_c; {S_\ell=2}\rangle_1,\\
 |  \frac{5}{2}_J\rangle &\equiv& |\frac{1}{2}_c; {S_\ell=2}\rangle_2.
\end{eqnarray}
Apparently, the spin-$5/2$ baryonic state  is the same with the one in Eq.~\eqref{eq:five_states_diquark}, while the two spin-$1/2$ and $3/2$ states will mix with each other respectively.  The mixing matrix is given as
\begin{eqnarray}
 |  \frac{1}{2}_J\rangle_{1'} &=& -\sqrt{\frac{1}{3}}  |L=1, \frac{1}{2}_J\rangle_{1}+  \sqrt{\frac{2}{3}}  |L=1, \frac{1}{2}_J \rangle_{2},\\
 |  \frac{1}{2}_J\rangle_{2'} &=& -\sqrt{\frac{2}{3}}  |L=1, \frac{1}{2}_J\rangle_{1}-  \sqrt{\frac{1}{3}}  |L=1, \frac{1}{2}_J \rangle_{2},
\end{eqnarray}
for the two spin-1/2 states and
\begin{eqnarray}
|\frac{3}{2}_J\rangle_{1'}= \sqrt{\frac{1}{6}} |L=1,\frac{3}{2}_J\rangle_{1} +  \sqrt{\frac{5}{6}} |L=1,\frac{3}{2}_J\rangle_{2},\\
|\frac{3}{2}_J\rangle_{2'}= \sqrt{\frac{5}{6}} |L=1,\frac{3}{2}_J\rangle_{1} -  \sqrt{\frac{1}{6}} |L=1,\frac{3}{2}_J\rangle_{2},
\end{eqnarray}
for the spin-$3/2$ baryons.

In the heavy quark limit, the amplitudes of $\Omega_c^0(X)\to\Xi_c^+(\Xi_c^{'+})K^-$ can be expressed as
\begin{eqnarray}
&&{\cal A}(\Omega_c(J,J_z)\to \Xi_c^{(')}(J',J'_z) K(L,L_z)) \nonumber\\
&=&\sum\langle\frac{1}{2},S_{cz};S_\ell,S_{\ell z}|J,J_z\rangle\langle\frac{1}{2},S_{cz};S'_\ell,S'_{\ell z}|J',J'_z\rangle\nonumber\\
&&\times\langle L,S'_{\ell};||{\cal H}_{eff}||S_\ell\rangle\langle L,L_z;S'_\ell,S'_{\ell z}|S_\ell,S_{\ell z}\rangle,
\label{eq:decay1}
\end{eqnarray}
where the quantum numbers $S_\ell$ and $S'_\ell$ are the  spin of the light degrees
of freedom in $\Omega_c^0(X)$ and $\Xi_c^+(\Xi_c^{'+})$ respectively, the quantum numbers $J$ and $J'$ are the total angular momentum of $\Omega_c^0(X)$ and $\Xi_c^+(\Xi_c^{'+})$ respectively.

The decay widths of $\Omega_c^0(X)\to\Xi_c^+(\Xi_c^{'+})K^-$ are proportional to Clebsch-Gordan coefficients
\begin{eqnarray}
\Gamma&\propto&(2S_\ell+1)(2J'+1)\left|\left\{\begin{array}{ccc}
                                        L & S'_\ell & S_\ell \\
                                        \frac{1}{2} & J & J'
                                      \end{array}\right\}\right|^2,
\label{eq:decay2}
\end{eqnarray}
where the product of Clebsch-Gordan coefficients are in terms of 6j symbols.

For $\Omega_c^0(X)\to\Xi_c^+K^-$, the quantum numbers are
\begin{eqnarray}
S_\ell'=0,\;\; S_\ell=(0,1,2),\;\; J'=\frac{1}{2},\;J=\left(\frac{1}{2},\frac{3}{2},\frac{5}{2}\right).
\end{eqnarray}
We find the following results:
\begin{itemize}

\item Due to the parity conservation, the decays can proceed through S-wave or D-wave.

\item Only the lowest-lying state, $ |\frac{1}{2}\rangle_{S_\ell=0}$ can decays into the $\Xi_c K$ in S-wave. The $ |\frac{1}{2}\rangle_{S_\ell=0}$ may mix with $ |\frac{1}{2}\rangle_{S_\ell=1}$ in QCD. However we expect that their low masses do not allow a large phase space. So the $1/2$ states will have not large decay widths.

\item The $ |\frac{3}{2}\rangle_{S_\ell=2}$ and  $ |\frac{5}{2}\rangle_{S_\ell=2}$ can decays into the $\Xi_c K$ through D-wave.  For the $ |\frac{5}{2}\rangle_{S_\ell=2}$, this is guaranteed by the angular momentum conservation, and while the heavy quark symmetry relates the decays of $ |\frac{3}{2}\rangle_{S_\ell=2}$.  Such amplitudes are also suppressed due to the phase space. Thus the total widths are expected to be  small again.

\item The breaking of heavy quark symmetry may induce small contributions to decay widths.

\end{itemize}

For the channel $\Omega_c^0(X)\to\Xi_c^{'+}K^-$, the related  quantum numbers of the initial and final states are
\begin{eqnarray}
S_\ell'=1,\;\; S_\ell=(0,1,2),\;\; J'=\frac{1}{2},\;J=(\frac{1}{2},\frac{3}{2},\frac{5}{2}).
\end{eqnarray}
The following remarks are given in order.
\begin{itemize}
\item The threshold of $\Xi_c^{'+}K^-$ is about 3069 MeV,  which prohibits   decays of the lower three baryons.

\item Decays of $\Omega(3090)$ and $\Omega(3119)$ into $\Xi_c^{'+}K^-$  have some phase space.

\item
From the 6j symbol, we find the    S-wave decay is through  $|1/2\rangle_{S_\ell=1}\to \Xi_c' K$.  But considering the threshold of $\Xi_c^{'+}K^-$ is about 3069 MeV,  this will not be kinematically allowed.

\item There are D-wave decay amplitudes for $|1/2\rangle_{S_\ell=1}\to \Xi_c' K$, $|3/2\rangle_{S_\ell=2}\to \Xi_c' K$, $|5/2\rangle_{S_\ell=2}\to \Xi_c' K$. However these contributions are not big since the phase space is limited.

\end{itemize}

Since both decays into $\Xi_c^+K^-$ and $\Xi_c^{'+}K^-$ are suppressed,
the narrowness of the five newly  observed $\Omega_c$ states   can be understood using heavy quark symmetry.

In the heavy-quark-light-diquark model,
the decay of $\Omega_c$ into $\Xi_c^+ K^-$ requests to tear the $ss$ diquark apart, and thus  the calculation of the width decay into $\Xi_c K$ is beyond   the quark-diquark scheme mainly used in this work.  A tool to estimate the decay width might be using the flavor SU(3) symmetry to relate to other charmed baryons, for instance $\Gamma(\Lambda_c(2595)) =(2.6\pm 0.6)$ MeV, $\Gamma(\Lambda_c(2625))  <0.97$ MeV~\cite{Olive:2016xmw},   $\Xi_c^+(2645)=(2.1\pm0.2)$ MeV, $\Xi_c^+(2790)=(8.9\pm1.0)$~\cite{Yelton:2016fqw}.  This can give us a hint that the corresponding $\Omega_c$ states might be narrow. However a conclusive result requests the classification  of the $\Lambda_c$ and $\Xi_c$ baryons and  a more comprehensive analysis to be published in future.

\section*{Conclusion}

In this work, we have  studied the  charmed and bottomed   baryons with two strange quarks in a quark-diquark model.  The two strange quarks lie in  S wave and thus their total spin  is 1. Within the heavy-quark-light-diquark framework, we calculate  the mass spectra  of the $S$ and $P$ wave orbitally excited states.  We find the   $\Omega_c^0 (2695)$ and $\Omega_c^0 (2770)$ fit well as the $S$ wave states of  charmed doubly strange baryons. There are five P-wave states.  The five newly  $\Omega_c^0$ resonances observed by the LHCb Collaboration, i.e. $\Omega_c^0(3000)$, $\Omega_c^0(3050)$, $\Omega_c^0(3066)$, $\Omega_c^0(3090)$, and $\Omega_c^0(3119)$, can be interpreted as the $P$ wave orbitally excited states of  charmed doubly strange baryons.  We have  analyzed their decays into the $\Xi_c K$ and $\Xi_c' K$ in the heavy quark effective theory. We find   decays of  the five new $\Omega_c$ states into the $\Xi_cK$ and $\Xi_c'K$ are suppressed by the heavy quark symmetry or the phase space. The narrowness of the five newly  observed $\Omega_c$ states   can be understood using heavy quark symmetry.

$$\textbf{Note added}$$

While this paper was submitted, there are studies of the masses or (and) decay properties of the newly  observed $\Omega_c^0(X)$ states using different approaches:
the QCD sum rules~\cite{Wang:2017zjw,Zhao:2017fov,Chen:2017gnu,Aliev:2017led,Agaev:2017lip}, heavy hadron chiral perturbation theory~\cite{Cheng:2017ove}, the chiral quark-soliton model~\cite{Kim:2017jpx},  and lattice QCD~\cite{Padmanath:2017lng},  the
constituent quark models and treatment as  pentaquarks~\cite{Huang:2017dwn,An:2017lwg}.

\section*{Acknowledgments}
This work was supported in part by the National Natural Science Foundation
of China under Grant No. 11575110, 11647163, 11655002, and by the Research Start-up Funding (R.L. Zhu) of
Nanjing Normal University,  by  Natural  Science Foundation of Shanghai under Grant  No. 15ZR1423100
 and No. 15DZ2272100,  by the Young Thousand Talents Plan,   and  by    Key Laboratory for Particle Physics, Astrophysics and Cosmology, Ministry of Education.


\begin{references}
\frenchspacing
%\cite{Aaij:2015tga}
\bibitem{Aaij:2015tga}
  R.~Aaij {\it et al.} (LHCb Collaboration),
  %``Observation of $J/\psi p$ Resonances Consistent with Pentaquark States in $\Lambda_b^0 \to J/\psi K^- p$ Decays,''
  Phys.\ Rev.\ Lett.\  {\bf 115}, 072001 (2015).
  %doi:10.1103/PhysRevLett.115.072001
  %[arXiv:1507.03414 [hep-ex]].
  %%CITATION = doi:10.1103/PhysRevLett.115.072001;%%
  %403 citations counted in INSPIRE as of 20 Jun 2017


%\cite{Aaij:2016ymb}
\bibitem{Aaij:2016ymb}
  R.~Aaij {\it et al.} (LHCb Collaboration),
  %``Evidence for exotic hadron contributions to $\Lambda_b^0 \to J/\psi p \pi^-$ decays,''
Phys.\ Rev.\ Lett.\  {\bf 117},  082003 (2016);
{\bf 117},109902(A) (2016);
{\bf 118}, 119901(A) (2017).
%doi:10.1103/PhysRevLett.118.119901, 10.1103/PhysRevLett.117.082003, 10.1103/PhysRevLett.117.109902
%[arXiv:1606.06999 [hep-ex]].
%%CITATION = doi:10.1103/PhysRevLett.118.119901, 10.1103/PhysRevLett.117.082003, 10.1103/PhysRevLett.117.109902;%%
  %40 citations counted in INSPIRE as of 09 Jun 2017



%\cite{Biagi:1984mu}
\bibitem{Biagi:1984mu}
  S.~F.~Biagi {\it et al.},
  %``Properties of the Charmed Strange Baryon A+ and Evidence for the Charmed Doubly Strange Baryon T0 at 2.74-GeV/c**2,''
  Z.\ Phys.\ C {\bf 28}, 175 (1985).
 % doi:10.1007/BF01575721
  %%CITATION = doi:10.1007/BF01575721;%%
  %163 citations counted in INSPIRE as of 31 Mar 2017


%\cite{Albrecht:1992xa}
\bibitem{Albrecht:1992xa}
  H.~Albrecht {\it et al.} (ARGUS Collaboration),
  %``Evidence for the production of the charmed, doubly strange baryon omega(c) in e+ e- annihilation,''
  Phys.\ Lett.\ B {\bf 288}, 367 (1992).
 % doi:10.1016/0370-2693(92)91116-Q
  %%CITATION = doi:10.1016/0370-2693(92)91116-Q;%%
  %51 citations counted in INSPIRE as of 31 Mar 2017


%\cite{Frabetti:1992bm}
\bibitem{Frabetti:1992bm}
  P.~L.~Frabetti {\it et al.} (E687 Collaboration),
  %``First evidence of omega(c)0 ---> omega- pi+,''
  Phys.\ Lett.\ B {\bf 300}, 190 (1993).
  %doi:10.1016/0370-2693(93)90769-E
  %%CITATION = doi:10.1016/0370-2693(93)90769-E;%%
  %66 citations counted in INSPIRE as of 31 Mar 2017


%\cite{Aubert:2006je}
\bibitem{Aubert:2006je}
  B.~Aubert {\it et al.} (BaBar Collaboration),
  %``Observation of an excited charm baryon Omega(C)* decaying to Omega(C)0 gamma,''
  Phys.\ Rev.\ Lett.\  {\bf 97}, 232001 (2006).
  %doi:10.1103/PhysRevLett.97.232001
  %[hep-ex/0608055].
  %%CITATION = doi:10.1103/PhysRevLett.97.232001;%%
  %102 citations counted in INSPIRE as of 31 Mar 2017


%\cite{Solovieva:2008fw}
\bibitem{Solovieva:2008fw}
  E.~Solovieva {\it et al.},
  %``Study of \Omega_c^0 and \Omega_c^{*0} Baryons at Belle,''
  Phys.\ Lett.\ B {\bf 672}, 1 (2009).
  %doi:10.1016/j.physletb.2008.12.062
  %[arXiv:0808.3677 [hep-ex]].
  %%CITATION = doi:10.1016/j.physletb.2008.12.062;%%
  %30 citations counted in INSPIRE as of 31 Mar 2017


%\cite{Aaij:2017nav}
\bibitem{Aaij:2017nav}
  R.~Aaij {\it et al.} (LHCb Collaboration),
  %``Observation of five new narrow $\Omega_c^0$ states decaying to $\Xi_c^+ K^-$,''
Phys.\ Rev.\ Lett.\  {\bf 118},  182001 (2017).
%doi:10.1103/PhysRevLett.118.182001
%[arXiv:1703.04639 [hep-ex]].
%%CITATION = doi:10.1103/PhysRevLett.118.182001;%%
  %23 citations counted in INSPIRE as of 09 Jun 2017


%\cite{Agaev:2017jyt}
\bibitem{Agaev:2017jyt}
  S.~S.~Agaev, K.~Azizi, and H.~Sundu,
  %``On the nature of the newly discovered $\Omega_c^{0}$ states,''
  arXiv:1703.07091 [hep-ph].
  %%CITATION = ARXIV:1703.07091;%%
  %4 citations counted in INSPIRE as of 31 Mar 2017


%\cite{Chen:2017sci}
\bibitem{Chen:2017sci}
  H.~X.~Chen, Q.~Mao, W.~Chen, A.~Hosaka, X.~Liu, and S.~L.~Zhu,
  %``Decay properties of $P$-wave charmed baryons from light-cone QCD sum rules,''
Phys.\ Rev.\ D {\bf 95}, 094008 (2017).
%doi:10.1103/PhysRevD.95.094008
%[arXiv:1703.07703 [hep-ph]].
%%CITATION = doi:10.1103/PhysRevD.95.094008;%%
  %12 citations counted in INSPIRE as of 09 Jun 2017


%\cite{Karliner:2017kfm}
\bibitem{Karliner:2017kfm}
  M.~Karliner and J.~L.~Rosner,
  %``Very narrow excited $\Omega_c$ baryons,''
  arXiv:1703.07774 [hep-ph].
  %%CITATION = ARXIV:1703.07774;%%
  %2 citations counted in INSPIRE as of 31 Mar 2017


%\cite{Wang:2017hej}
\bibitem{Wang:2017hej}
  K.~L.~Wang, L.~Y.~Xiao, X.~H.~Zhong, and Q.~Zhao,
  %``Understanding the newly observed $\Omega_c$ states through their decays,''
  arXiv:1703.09130 [hep-ph].
  %%CITATION = ARXIV:1703.09130;%%


%\cite{Yang:2017rpg}
\bibitem{Yang:2017rpg}
  G.~Yang and J.~Ping,
  %``The structure of pentaquarks $\Omega_c^0$ in the chiral quark model,''
  arXiv:1703.08845 [hep-ph].
  %%CITATION = ARXIV:1703.08845;%%

%\cite{GellMann:1964nj}
\bibitem{GellMann:1964nj}
  M.~Gell-Mann,
%``A Schematic Model of Baryons and Mesons,''
Phys.\ Lett.\  {\bf 8}, 214 (1964).
%doi:10.1016/S0031-9163(64)92001-3
%%CITATION = doi:10.1016/S0031-9163(64)92001-3;%%
  %2536 citations counted in INSPIRE as of 17 Jun 2017



%\cite{Brodsky:2014xia}
\bibitem{Brodsky:2014xia}
  S.~J.~Brodsky, D.~S.~Hwang, and R.~F.~Lebed,
  %``Dynamical Picture for the Formation and Decay of the Exotic XYZ Mesons,''
Phys.\ Rev.\ Lett.\  {\bf 113}, no. 11, 112001 (2014).
%doi:10.1103/PhysRevLett.113.112001
%[arXiv:1406.7281 [hep-ph]].
%%CITATION = doi:10.1103/PhysRevLett.113.112001;%%
%63 citations counted in INSPIRE as of 01 Apr 2017

%\cite{Jaffe:2003sg}
\bibitem{Jaffe:2003sg}
  R.~L.~Jaffe and F.~Wilczek,
  %``Diquarks and exotic spectroscopy,''
  Phys.\ Rev.\ Lett.\  {\bf 91}, 232003 (2003).
  %doi:10.1103/PhysRevLett.91.232003
  %[hep-ph/0307341].
  %%CITATION = doi:10.1103/PhysRevLett.91.232003;%%
  %789 citations counted in INSPIRE as of 31 Mar 2017

%\cite{Jaffe:2004ph}
\bibitem{Jaffe:2004ph}
  R.~L.~Jaffe,
  %``Exotica,''
  Phys.\ Rept.\  {\bf 409}, 1 (2005).
  %doi:10.1016/j.physrep.2004.11.005
  %[hep-ph/0409065].
  %%CITATION = doi:10.1016/j.physrep.2004.11.005;%%
  %391 citations counted in INSPIRE as of 20 Jun 2017

%\cite{Maiani:2004vq}
\bibitem{Maiani:2004vq}
  L.~Maiani, F.~Piccinini, A.~D.~Polosa, and V.~Riquer,
  %``Diquark-antidiquarks with hidden or open charm and the nature of X(3872),''
  Phys.\ Rev.\ D {\bf 71}, 014028 (2005).
  %doi:10.1103/PhysRevD.71.014028
 % [hep-ph/0412098].
  %%CITATION = doi:10.1103/PhysRevD.71.014028;%%
  %562 citations counted in INSPIRE as of 20 Jun 2017

%\cite{Ali:2011ug}
\bibitem{Ali:2011ug}
  A.~Ali, C.~Hambrock, and W.~Wang,
  %``Tetraquark Interpretation of the Charged Bottomonium-like states $Z_b^{+-}(10610)$ and $Z_b^{+-}(10650)$ and Implications,''
  Phys.\ Rev.\ D {\bf 85}, 054011 (2012).
  %doi:10.1103/PhysRevD.85.054011
  %[arXiv:1110.1333 [hep-ph]].
  %%CITATION = doi:10.1103/PhysRevD.85.054011;%%
  %72 citations counted in INSPIRE as of 20 Jun 2017

\bibitem{Hambrock:2011qga}
  C.~Hambrock, DESY-THESIS-2011-012.

%\cite{He:2016xvd}
\bibitem{He:2016xvd}
  X.~G.~He, W.~Wang, and R.~L.~Zhu,
  %``Production of Charmed Tetraquarks from $B_c$ and $B$ decays,''
  J.\ Phys.\ G {\bf 44},  014003 (2017).
  %doi:10.1088/0954-3899/44/1/014003,
  %[arXiv:1606.00097 [hep-ph]].
  %%CITATION = doi:10.1088/0954-3899/44/1/014003, 10.1088/0022-3727/44/27/274003;%%
  %5 citations counted in INSPIRE as of 20 Jun 2017

%\cite{Ali:2009es}
\bibitem{Ali:2009es}
  A.~Ali, C.~Hambrock, and M.~J.~Aslam,
  %``A Tetraquark interpretation of the BELLE data on the anomalous Upsilon(1S) pi+pi- and Upsilon(2S) pi+pi- production near the Upsilon(5S) resonance,''
  Phys.\ Rev.\ Lett.\  {\bf 104}, 162001 (2010);{\bf 107}, 049903(E) (2011).
  %Erratum: [Phys.\ Rev.\ Lett.\  {\bf 107}, 049903 (2011)]
 % doi:10.1103/PhysRevLett.104.162001, 10.1103/PhysRevLett.107.049903
  %[arXiv:0912.5016 [hep-ph]].
  %%CITATION = doi:10.1103/PhysRevLett.104.162001, 10.1103/PhysRevLett.107.049903;%%
  %62 citations counted in INSPIRE as of 20 Jun 2017

%\cite{Zhu:2015bba}
\bibitem{Zhu:2015bba}
  R.~Zhu and C.~F.~Qiao,
  %``Pentaquark states in a diquark-triquark model,''
  Phys.\ Lett.\ B {\bf 756}, 259 (2016).
  %doi:10.1016/j.physletb.2016.03.022
  %[arXiv:1510.08693 [hep-ph]].
  %%CITATION = doi:10.1016/j.physletb.2016.03.022;%%
  %36 citations counted in INSPIRE as of 20 Jun 2017


%\cite{Wang:2016tsi}
\bibitem{Wang:2016tsi}
  W.~Wang and R.~Zhu,
  %``Can $X(5568)$ be a tetraquark state?,''
  Chin.\ Phys.\ C {\bf 40},  093101 (2016).
  %doi:10.1088/1674-1137/40/9/093101
  %[arXiv:1602.08806 [hep-ph]].
  %%CITATION = doi:10.1088/1674-1137/40/9/093101;%%
  %42 citations counted in INSPIRE as of 20 Jun 2017


%\cite{Zhu:2016arf}
\bibitem{Zhu:2016arf}
  R.~Zhu,
  %``Hidden charm octet tetraquarks from a diquark-antidiquark model,''
  Phys.\ Rev.\ D {\bf 94},  054009 (2016).
  %doi:10.1103/PhysRevD.94.054009
  %[arXiv:1607.02799 [hep-ph]].
  %%CITATION = doi:10.1103/PhysRevD.94.054009;%%
  %8 citations counted in INSPIRE as of 20 Jun 2017

%\cite{Olive:2016xmw}
\bibitem{Olive:2016xmw}
  C.~Patrignani {\it et al.} (Particle Data Group),
  %``Review of Particle Physics,''
  Chin.\ Phys.\ C {\bf 40},  100001 (2016).
  %doi:10.1088/1674-1137/40/10/100001
  %%CITATION = doi:10.1088/1674-1137/40/10/100001;%%
  %1018 citations counted in INSPIRE as of 20 Jun 2017

%\cite{JimenezTejero:2009vq}
\bibitem{JimenezTejero:2009vq}
  C.~E.~Jimenez-Tejero, A.~Ramos, and I.~Vidana,
%``Dynamically generated open charmed baryons beyond the zero range approximation,''
Phys.\ Rev.\ C {\bf 80}, 055206 (2009).
%doi:10.1103/PhysRevC.80.055206
%[arXiv:0907.5316 [hep-ph]].
%%CITATION = doi:10.1103/PhysRevC.80.055206;%%
  %47 citations counted in INSPIRE as of 18 Jun 2017

%\cite{Romanets:2012hm}
\bibitem{Romanets:2012hm}
  O.~Romanets, L.~Tolos, C.~Garcia-Recio, J.~Nieves, L.~L.~Salcedo, and R.~G.~E.~Timmermans,
%``Charmed and strange baryon resonances with heavy-quark spin symmetry,''
Phys.\ Rev.\ D {\bf 85}, 114032 (2012).
%doi:10.1103/PhysRevD.85.114032
%[arXiv:1202.2239 [hep-ph]].
%%CITATION = doi:10.1103/PhysRevD.85.114032;%%
  %61 citations counted in INSPIRE as of 18 Jun 2017




%\cite{Shah:2016nxi}
\bibitem{Shah:2016nxi}
  Z.~Shah, K.~Thakkar, A.~K.~Rai, and P.~C.~Vinodkumar,
  %``Mass spectra and Regge trajectories of $\Lambda_{c}^{+}$, $\Sigma_{c}^{0}$, $\Xi_{c}^{0}$ and $\Omega_{c}^{0}$ baryons,''
  Chin.\ Phys.\ C {\bf 40},  123102 (2016).
  %doi:10.1088/1674-1137/40/12/123102
  %[arXiv:1609.08464 [nucl-th]].
  %%CITATION = doi:10.1088/1674-1137/40/12/123102;%%
  %7 citations counted in INSPIRE as of 31 Mar 2017


%\cite{Ebert:2011kk}
\bibitem{Ebert:2011kk}
  D.~Ebert, R.~N.~Faustov, and V.~O.~Galkin,
  %``Spectroscopy and Regge trajectories of heavy baryons in the relativistic quark-diquark picture,''
  Phys.\ Rev.\ D {\bf 84}, 014025 (2011).
  %doi:10.1103/PhysRevD.84.014025
  %[arXiv:1105.0583 [hep-ph]].
  %%CITATION = doi:10.1103/PhysRevD.84.014025;%%
  %53 citations counted in INSPIRE as of 31 Mar 2017


%\cite{Roberts:2007ni}
\bibitem{Roberts:2007ni}
  W.~Roberts and M.~Pervin,
  %``Heavy baryons in a quark model,''
  Int.\ J.\ Mod.\ Phys.\ A {\bf 23}, 2817 (2008).
  %doi:10.1142/S0217751X08041219
  %[arXiv:0711.2492 [nucl-th]].
  %%CITATION = doi:10.1142/S0217751X08041219;%%
  %181 citations counted in INSPIRE as of 31 Mar 2017


%\cite{Yoshida:2015tia}
\bibitem{Yoshida:2015tia}
  T.~Yoshida, E.~Hiyama, A.~Hosaka, M.~Oka, and K.~Sadato,
  %``Spectrum of heavy baryons in the quark model,''
  Phys.\ Rev.\ D {\bf 92},  114029 (2015).
  %doi:10.1103/PhysRevD.92.114029
  %[arXiv:1510.01067 [hep-ph]].
  %%CITATION = doi:10.1103/PhysRevD.92.114029;%%
  %28 citations counted in INSPIRE as of 31 Mar 2017


%\cite{Chen:2015kpa}
\bibitem{Chen:2015kpa}
  H.~X.~Chen, W.~Chen, Q.~Mao, A.~Hosaka, X.~Liu, and S.~L.~Zhu,
  %``P-wave charmed baryons from QCD sum rules,''
  Phys.\ Rev.\ D {\bf 91}, no. 5, 054034 (2015).
  %doi:10.1103/PhysRevD.91.054034
  %[arXiv:1502.01103 [hep-ph]].
  %%CITATION = doi:10.1103/PhysRevD.91.054034;%%
  %19 citations counted in INSPIRE as of 31 Mar 2017


%\cite{Valcarce:2008dr}
\bibitem{Valcarce:2008dr}
  A.~Valcarce, H.~Garcilazo, and J.~Vijande,
  %``Towards an understanding of heavy baryon spectroscopy,''
  Eur.\ Phys.\ J.\ A {\bf 37}, 217 (2008).
  %doi:10.1140/epja/i2008-10616-4
  %[arXiv:0807.2973 [hep-ph]].
  %%CITATION = doi:10.1140/epja/i2008-10616-4;%%
  %59 citations counted in INSPIRE as of 31 Mar 2017


%\cite{Yamaguchi:2014era}
\bibitem{Yamaguchi:2014era}
  Y.~Yamaguchi, S.~Ohkoda, A.~Hosaka, T.~Hyodo, and S.~Yasui,
  %``Heavy quark symmetry in multihadron systems,''
  Phys.\ Rev.\ D {\bf 91}, 034034 (2015).
 % doi:10.1103/PhysRevD.91.034034
  %[arXiv:1402.5222 [hep-ph]].
  %%CITATION = doi:10.1103/PhysRevD.91.034034;%%
  %16 citations counted in INSPIRE as of 31 Mar 2017


%\cite{Bali:2015lka}
\bibitem{Bali:2015lka}
  P.~Prez-Rubio, S.~Collins, and G.~S.~Bali,
  %``Charmed baryon spectroscopy and light flavor symmetry from lattice QCD,''
  Phys.\ Rev.\ D {\bf 92}, 034504 (2015).
  %doi:10.1103/PhysRevD.92.034504
  %[arXiv:1503.08440 [hep-lat]].
  %%CITATION = doi:10.1103/PhysRevD.92.034504;%%
  %26 citations counted in INSPIRE as of 31 Mar 2017

%\cite{Yelton:2016fqw}
\bibitem{Yelton:2016fqw}
  J.~Yelton {\it et al.} (Belle Collaboration),
  %``Study of Excited $\Xi_c$ States Decaying into $\Xi_c^0$ and
%$\Xi_c^+$ Baryons,''
  Phys.\ Rev.\ D {\bf 94},  052011 (2016).
  %doi:10.1103/PhysRevD.94.052011
  %[arXiv:1607.07123 [hep-ex]].
  %%CITATION = doi:10.1103/PhysRevD.94.052011;%%
  %13 citations counted in INSPIRE as of 28 Jun 2017

%\cite{Wang:2017zjw}
\bibitem{Wang:2017zjw}
  Z.~G.~Wang,
  %``Analysis of $\Omega _c(3000)$ , $\Omega _c(3050)$ , $\Omega _c(3066)$ , $\Omega _c(3090)$ and $\Omega _c(3119)$ with QCD sum rules,''
Eur.\ Phys.\ J.\ C {\bf 77},  325 (2017).
%doi:10.1140/epjc/s10052-017-4895-5
%[arXiv:1704.01854 [hep-ph]].
%%CITATION = doi:10.1140/epjc/s10052-017-4895-5;%%
  %8 citations counted in INSPIRE as of 09 Jun 2017

  %\cite{Zhao:2017fov}
\bibitem{Zhao:2017fov}
  Z.~Zhao, D.~D.~Ye, and A.~Zhang,
  %``Hadronic decay properties of newly observed $\Omega_c$ baryons,''
arXiv:1704.02688 [hep-ph].
%%CITATION = ARXIV:1704.02688;%%
  %4 citations counted in INSPIRE as of 09 Jun 2017

  %\cite{Chen:2017gnu}
\bibitem{Chen:2017gnu}
  B.~Chen and X.~Liu,
  %``Six $\Omega_c^0$ states discovered by LHCb as new members of $1P$ and $2S$ charmed baryons,''
arXiv:1704.02583 [hep-ph].
%%CITATION = ARXIV:1704.02583;%%
  %5 citations counted in INSPIRE as of 09 Jun 2017

%\cite{Aliev:2017led}
\bibitem{Aliev:2017led}
  T.~M.~Aliev, S.~Bilmis, and M.~Savci,
  %``Are the new excited $\Omega_c$ baryons negative parity states?,''
arXiv:1704.03439 [hep-ph].
%%CITATION = ARXIV:1704.03439;%%
  %6 citations counted in INSPIRE as of 09 Jun 2017

%\cite{Agaev:2017lip}
\bibitem{Agaev:2017lip}
S.~S.~Agaev, K.~Azizi, and H.~Sundu,
%``Interpretation of the new $\Omega_c^{0}$ states via their mass and width,''
arXiv:1704.04928 [hep-ph].
%%CITATION = ARXIV:1704.04928;%%
  %3 citations counted in INSPIRE as of 09 Jun 2017

%\cite{Cheng:2017ove}
\bibitem{Cheng:2017ove}
  H.~Y.~Cheng and C.~W.~Chiang,
  %``Quantum numbers of $\Omega_c$ states and other charmed baryons,''
Phys.\ Rev.\ D {\bf 95}, 094018 (2017).
%doi:10.1103/PhysRevD.95.094018
%[arXiv:1704.00396 [hep-ph]].
%%CITATION = doi:10.1103/PhysRevD.95.094018;%%
  %8 citations counted in INSPIRE as of 09 Jun 2017

%\cite{Kim:2017jpx}
\bibitem{Kim:2017jpx}
  H.~C.~Kim, M.~V.~Polyakov, and M.~Praszalowicz,
%``On a possibility of charmed exotica,''
arXiv:1704.04082 [hep-ph].
%%CITATION = ARXIV:1704.04082;%%
  %5 citations counted in INSPIRE as of 09 Jun 2017


 %\cite{Padmanath:2017lng}
\bibitem{Padmanath:2017lng}
  M.~Padmanath and N.~Mathur,
%``Quantum numbers of recently discovered $\Omega^{0}_{c}$ baryons from lattice QCD,''
arXiv:1704.00259 [hep-ph].
%%CITATION = ARXIV:1704.00259;%%
  %10 citations counted in INSPIRE as of 09 Jun 2017

%\cite{Huang:2017dwn}
\bibitem{Huang:2017dwn}
  H.~Huang, J.~Ping, and F.~Wang,
%``Investigating the excited $\Omega^{0}_{c}$ states through $\Xi_{c}K$ and $\Xi^{'}_{c}K$ decay channels,''
arXiv:1704.01421 [hep-ph].
%%CITATION = ARXIV:1704.01421;%%
  %8 citations counted in INSPIRE as of 09 Jun 2017

%\cite{An:2017lwg}
\bibitem{An:2017lwg}
C.~S.~An and H.~Chen,
%``The observed $\Omega_{c}^{0}$ resonances as pentaquark states,''
arXiv:1705.08571 [hep-ph].
%%CITATION = ARXIV:1705.08571;%%



\end{references}
\end{document}